\documentclass{article}
\usepackage{spconf,graphicx}
\usepackage{booktabs}
\usepackage{amsmath,amssymb,amsfonts}
\usepackage{bm}
\usepackage{algorithmic}
\usepackage{algorithm}
\usepackage{array}
\usepackage{stfloats}
\usepackage[acronym]{glossaries}
\usepackage{microtype}
\usepackage{url}
\makeglossaries

\ninept

\renewcommand{\vec}[1]{\bm{#1}}
\newcommand{\mat}[1]{\bm{#1}}

\newcommand{\diag}[1]{\operatorname{diag}\left[ #1 \right]} 

\def\ie{i.\,e.\,,\ }
\def\eg{e.\,g.\,,\ }

\newacronym{hello}{hw}{hello world}
\newacronym{doa}{DoA}{direction of arrival}
\newacronym{rir}{RIR}{room impulse response}
\newacronym{srir}{SRIR}{spatial room impulse response}

\newacronym{sh}{SH}{spherical harmonic}
\newacronym{shd}{SHD}{spherical harmonic domain}
\newacronym{sht}{SHT}{spherical harmonic transform}
\newacronym{isht}{iSHT}{inverse spherical harmonic transform}

\newacronym{hoa}{HOA}{Higher-Order Ambisonics}
\newacronym{dirac}{DirAC}{Directional Audio Coding}
\newacronym{hodirac}{HO-DirAC}{Higher-Order Directional Audio Coding}
\newacronym{tc}{TC}{transport channel}

\newacronym{anova}{ANOVA}{analysis of variance}

\newcommand{\symmat}[1]{{\mbox{\boldmath $#1$}}}
\renewcommand{\vec}[1]{\bm{#1}}

\title{Perceptually-Motivated Spatial Audio Codec for \\ Higher-Order Ambisonics Compression} %
\name{Christoph Hold\textsuperscript{1}, Leo McCormack\textsuperscript{1}, Archontis Politis\textsuperscript{2}, Ville Pulkki\textsuperscript{1}}%
\address{\textsuperscript{1}Acoustics Lab, Dept. Information and Communications Eng., Aalto University, Espoo, Finland\\ %
\textsuperscript{2}Faculty of Information Technology and Communication Sciences, Tampere University, Finland}
\begin{document}
\maketitle
\begin{abstract}
Scene-based spatial audio formats, such as Ambisonics, are playback system agnostic and may therefore be favoured for delivering immersive audio experiences to a wide range of (potentially unknown) devices. The number of channels required to deliver high spatial resolution Ambisonic audio, however, can be prohibitive for low-bandwidth applications.
Therefore, this paper proposes a compression codec, which is based upon the parametric higher-order Directional Audio Coding (HO-DirAC) model. The encoder downmixes the higher-order Ambisonic (HOA) input audio into a reduced number of signals, which are accompanied by perceptually-motivated scene parameters. The downmixed audio is coded using a perceptual audio coder, whereas the parameters are grouped into perceptual bands, quantized, and downsampled. 
On the decoder side, low Ambisonic orders are fully recovered. Not fully recoverable HOA components are synthesized according to the parameters.
The results of a listening test indicate that the proposed parametric spatial audio codec can improve the adopted perceptual audio coder, especially at low to medium-high bitrates, when applied to fifth-order HOA signals.
\end{abstract}
\section{Introduction}
\label{sec:intro}

Consumption of spatial audio is growing within the consumer space, with audio streaming companies now delivering multi-channel immersive audio experiences to their user base. Much of these immersive audio reproduction formats, however, are based on discrete/fixed loudspeaker channel layouts, also utilized for the more commonly consumed (optionally head-tracked) binauralized version for headphones playback. This is despite alternative formats, such as (higher-order) Ambisonics~\cite{gerzon1973periphony,zotter2019ambisonics}, being available, which has no such reproduction layout dependency; thus, offering greater playback flexibility. 
The primary reason for the lack of widespread adoption of Ambisonics is that higher-orders are required to deliver a high spatial resolution rendition of the scene \cite{bertet2013investigation,avni2013spatial}. Here, the number of audio channels scales quadratically with the encoding order, which means that the bandwidth required to transmit HOA scenes can be prohibitive for many applications; especially for streaming content. Audio compression codecs have, however, been explored for reducing Ambisonic transmission requirements. 
One open-source example is Opus \cite{valin2012definition}, which has been investigated for the task of HOA compression in \cite{rudzki2019perceptual,rudzki2019auditory} using various different bitrates, where lower bitrates are shown to reduce the perceived quality of the reproduced scene. Another recent study comparing the Opus Channel Mapping Family (CMF) found that beamformer based audio coding yields higher quality on average than spherical harmonic (SH) domain signal based~\cite{lee2023context}. An alternative coding is described in MPEG-H~\cite{herre2015mpeg} and proposed variations such as \cite{Zamani2017,Namazi2022,Xu2021}, separating HOA into foreground and background streams. However, owing to the closed-source nature of its development, the performance and various implementation details of MPEG-H encoders remain largely unknown to the wider scientific community. Other currently developed codecs (\eg IVAS \cite{ivas}) are limited to maximally third-order transmission, which may not be sufficient for large loudspeaker setups, or for avoiding coloration in binaural reproduction~\cite{Ahrens2019}.

Ambisonics are an orthogonal basis representation and, hence, no prior channel redundancy assumptions are met.
However, a single far field source can be expressed sufficiently as a single audio signal and its corresponding direction of arrival (DoA); hence its Ambisonic representation is not only limited, but also redundant across the channels.
Scene adaptive parameters can, therefore, leverage further compression gain, otherwise not available on purely the signal level.
Thus, another approach for compressing Ambisonic scenes is to adopt a parametric sound-field model, such as the Directional Audio Coding (DirAC) \cite{pulkki2007spatial} model, or its higher-order variant (HO-DirAC) \cite{hold2024compression}. The latter operates by segregating the sound-field into directionally constrained sectors, which are parameterised independently. These sector parameters are stored or transmitted, along with a reduced (down-mixed) number of audio channels compared to the original HOA audio.
Such an approach may be viewed as a possible extension to perceptual coders, with the parameterization also available for upmixing \cite{hold2023optimizing} and enabling certain sound-field modifications and spatial audio effects \cite{mccormack2021parametric}. Furthermore, it enables optimizations and trade-offs between audio and parameter datarate, which can be adapted to specific scenarios.
However, the compression performance achievable through extending perceptual audio coders with the HO-DirAC model, and how this might compare to existing multi-channel audio coders, has not yet been formally reported.

In this paper, the HO-DirAC architecture described in \cite{hold2024compression,hold2023optimizing} is adopted, and used to build a complete spatial audio codec. The codec applies perceptual audio coding (Opus) on the downmixed transport channels, and conducts grouping, quantization, and downsampling of the estimated HO-DirAC parameters. This combination is then used to reconstruct the HOA signals on the receiving end. By leveraging an understanding of spatial audio perception,  we believe that such a system represents an intuitive means of reducing bitrates for HOA scene transmission, which could provide a natural extension for already established perceptual audio codecs.

\section{Background}
The proposed codec operates in the time-frequency domain using the transform described in \cite{vilkamo2017time} ($t, f$ indices dropped for brevity).
We then aim to find a subset of $J$ beamformer signals, which represents the HOA input $\symmat{\chi}$ in terms of signal amplitude or energy~\cite{Hold2021a}.
Therefore, the encoder steers $J$ beamformers using a grid that is uniform or (if an appropriate uniform spherical sampling grid is unknown) near-uniform. %
For each sector $s \in [1, \ldots, J]$, applying the beamforming matrix $\mathbf{A}$ extracts audio transport channels (TCs) $\vec{x}$ corresponding to the local sound field pressure, and matrix $\mathbf{A}_{xyz}$ obtains the local sound field velocity components \cite{politis2015sector}.
This leads to the estimation of a time- and frequency-dependent signal and parameter vector as 
\begin{equation}
    \mathcal{A}[\mathbf{A}\symmat{\chi}, \mathbf{A}_{xyz}\symmat{\chi}] = [x_1, \Omega_1,  E_1, \psi_1, ..., x_J, \Omega_J,  E_J, \psi_J],
\end{equation}
where $\mathcal{A}[.]$ denotes performing the directional analysis described in~\cite{politis2015sector}; with $x_s$ being the audio TC per sector, accompanied by the sector energies $E_s$, the DoAs $\mathbf{\Omega} = [\Omega_1,...,\Omega_J]$ and diffuseness values ${\bm{\psi}} =[\psi_1,...,\psi_J]$; with the latter ranging from $[0, 1]$, where a single impinging plane-wave corresponds to $\psi_s=0$.
Both $\Omega, \psi$ are extracted from the perceptually-relevant active intensity vector.
These parameter estimates may also be further post-processed, \eg the diffuseness was filtered with a short third order median filter.

At the decoder, we aim to obtain the HOA signals $\tilde{\symmat{\chi}}$, based on a re-encoding strategy applied to $\vec{x}$.
Due to the structure of SHs, low orders can be recovered for $L < J$ using, for example, a truncated pseudo inverse $\check{\mathbf{B}} = \lfloor\mathbf{A}\rfloor^\dagger$; where $L$ is the number of SH components~\cite{hold2024compression}.
Higher orders, however, are not fully recoverable solely from the TCs, but we may instead synthesize $\hat{\symmat{\chi}}$ based on the HO-DirAC sound field model and the extracted sound field parameterization. Under the plane-wave assumption, we can use the SH matrix $\mat{Y}$ \cite{rafaely2015fundamentals}, and a factor $\beta_A$ (which is determined based on $\mat{A}$) as described in \cite{Hold2021a,Hold2021b}, which ensures reconstruction properties, as
\begin{equation}
    \hat{\symmat{\chi}} = \hat{\mathbf{B}} \diag{\bm{\psi}} \vec{x} + \beta_A \mathbf{Y}(\mathbf{\Omega}) \diag{1-\bm{\psi}} \vec{x}.
\end{equation}
This model may also be expressed compactly as a single matrix $\mathbf{M}$ %
\begin{equation}
    \hat{\symmat{\chi}} = \diag{\vec{g}} \mathbf{M} \vec{x},
\end{equation}
where order dependent gain factors $\vec{g}$ are introduced, developed in~\cite{hold2023optimizing}, utilizing the input parameterization to apply a model-based post-processing operation, which can adapt the codec output to match the original spatial covariance of the input.
We have found that matching the energy per order represents a simple, yet effective, strategy for mitigating coding shortcomings.

\begin{figure}[t]
    \centering
    \includegraphics[scale=0.35,trim={0cm 0cm 0cm 0cm},clip]{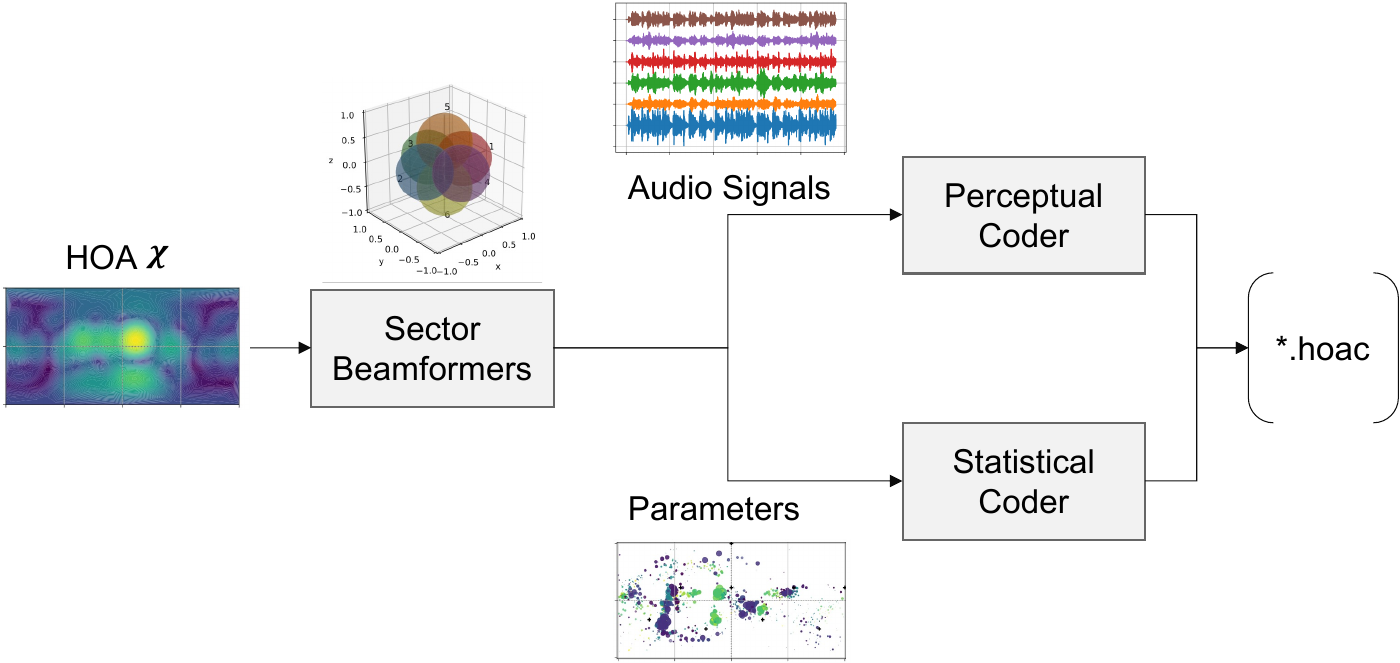}
    \\
    \vspace{1cm}
    \includegraphics[scale=0.35,trim={0cm 0cm 0cm 0cm},clip]{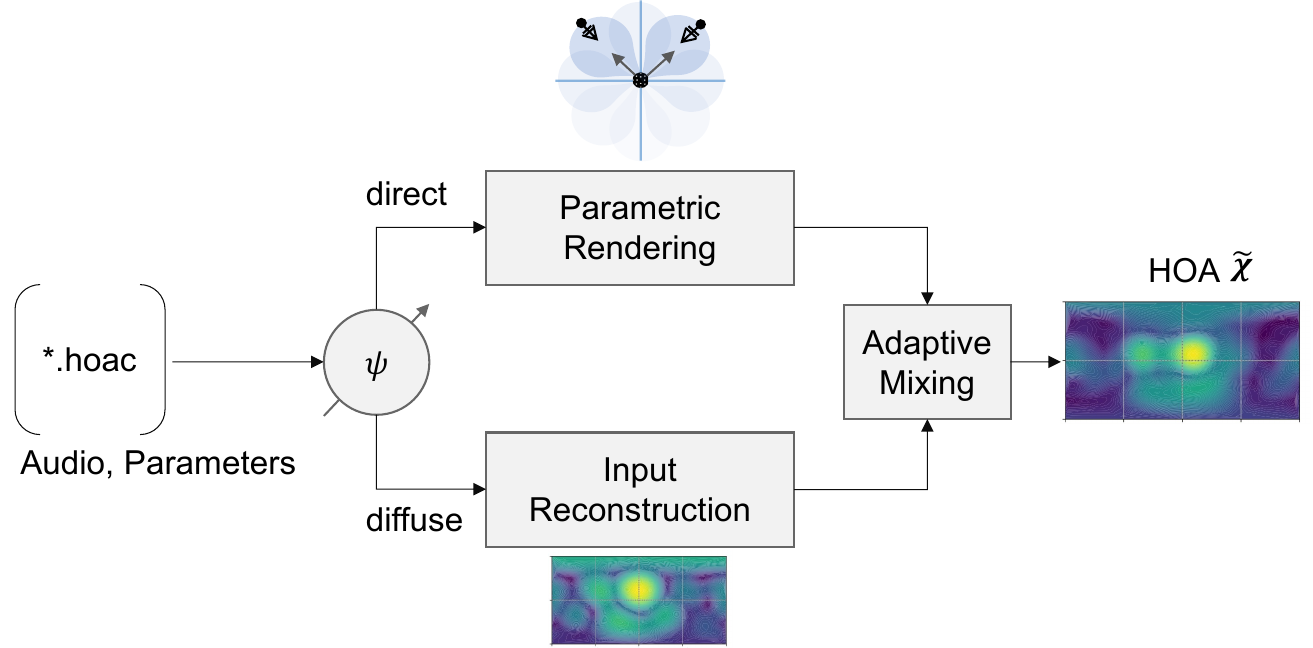}
    \caption{Overview of the proposed encoder (top) and decoder (bottom).}
    \label{fig:flow}
    \vspace{-0.5cm}
\end{figure}

\section{Proposed codec}
The proposed codec structure is illustrated in Fig.\,\ref{fig:flow}.
In the following, the adopted parameter quantization ideas are described.
First, the $133$ frequency band parameters are grouped into $16$ perceptual bands, corresponding approximately to the Bark scale \cite{zwicker1961subdivision}.
Within each band, the intensity vectors (before DoA and diffuseness estimation) are C-weighted, forming a perceptually-motivated average.
The resulting $16$ DoAs are quantized by finding the closest point within a uniformly spaced directional grid of $756$ points ($6.69$ degrees angular distance).
The diffuseness values are discretized to $8$ bins, which are non-uniformly spaced (with $\exp$ factor $1.25$), \ie finer resolution at small values. Both indices are then written to the parameter stream.

Note that additional grid directions, \eg placed on the horizontal plane, or other perceptually informed optimizations could be added at this stage.
Furthermore, adaptive scaling of the grid density, \eg based on energy and diffuseness of the DoA estimate, was also explored by the authors. Here, the quantization directional grid was collapsed when the estimated energy ($E_s$) was low or the 
diffuseness~($\psi_s$) was high.
However, these adaptive grid optimizations would require a dedicated evaluation, and were therefore omitted to favour a simpler presentation.
Instead, we noticed that high diffuseness estimates are linked to unreliable DoA estimates, and that the rendering would in any case shift towards the (non-DoA informed) diffuse-stream \cite{hold2024compression}. Therefore, we adopted a simpler coding optimisation by writing the DoA index as zero, during periods of very high diffuseness~(here $>0.95$, other theoretical limits may also be used~\cite{hold2024compression}). 
This was found to work well together with the subsequent run-length encoding Burrows–Wheeler transform~\cite{burrows1994block} (block sorting compression) preceding the Huffman-coding~\cite{huffman1952method}, which represents multiple consecutively identical values with high data efficiency.

Finally, we explored frequency-dependent down-sampling. Here the assumption is that low frequency estimates change less rapidly and can, therefore, be represented coarser in time.
We implemented this frequency-dependent down-sampling by simply eliminating time bins uniformly. %
The decoder is informed of this down-sampling and repeats the entries accordingly, before temporally smoothing the DoA; therefore, the effective rendering resolution also becomes higher than the otherwise seemingly sparse $756$-point quantization grid.
The parameters were then compressed using the \texttt{bzip2} algorithm, which supports incremental compression and decompression. An informal comparison with the \texttt{lzma} algorithm (which supports the same interface) showed a small advantage for \texttt{bzip2}.%

The perceptual audio coders adopted for the proposed codec are built around Opus \cite{valin2012definition}, which was selected due to its open, easily adaptable, and available source code.
We highlight that in principle any other perceptual audio coder could be adopted here, since the proposed framework operates independently of the audio coder.
The Opus family can be utilized in different modes, with and without stereo coding. For this study, we implemented a mode with independent channel coding, according to CMF\,255. This independent channel mode is used for the audio transport channels $\vec{x}$.
For the remaining implementation, we limited amplification in $\vec{g}$ to $6\,\mathrm{dB}$, estimated the parameterization utilizing $8$ frames of $128$ time-domain samples ($\mathrm{fs}=48\,\mathrm{kHz}$), and $\mathbf{M}$ was finally non-recursively smoothed with $2/3$ of the current and $+$ $1/3$ of the previous solution.

\begin{figure}
    \centering
    \includegraphics{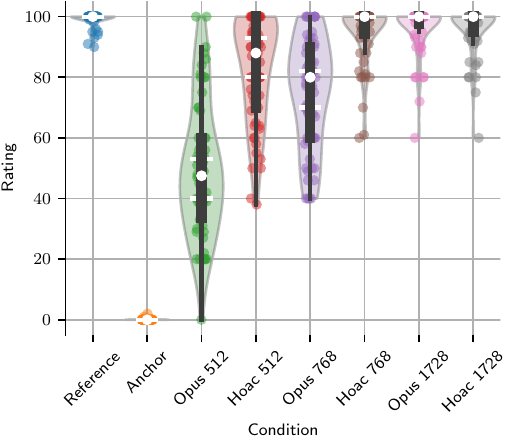}
    \caption{The medians are shown as white circles with their 95\% confidence intervals as white dashes, the box-plot with inter-quartile range in black, and the individual data points as colored dots.}
    \label{fig:results-TotalVio}
\end{figure}

\section{Experimental Design}
There are two CMFs in Opus that are specifically intended for HOA compression.
CMF\,2 runs the perceptual coders directly on the HOA channels.
However, this practice can easily corrupt the delicate inter-channel relations of the SHs, which can lead to noticeable degradations in perceptual quality, as previously identified in \cite{rudzki2019perceptual,lee2023context}.
Therefore, CMF\,3 is preferred, since it first steers a minimal grid of beamformers ($J=L$) and codes the static beamformer output signals; thus, %
an implementation of
Opus CMF\,3 represents our direct comparison/baseline codec.
The proposed HOA codec (Hoac) instead constructs the beamformers and reconstruction described in~\cite{Hold2021b,hold2024compression} (setting $J<L$), and adopts Opus CMF\,255 for coding the TCs, while performing the additional parameterization of the scene. 

The current intention is to investigate how the proposed additional scene parameterization layer (and reduced TCs) affects the overall coding (as opposed to evaluating the core-coder). %
Previous studies and our own informal testing suggest that Opus CMF\,3 may reach perceptual transparency at around $48\,\mathrm{kbps}$ (kilobits per second) per channel, when operating in variable bitrate (VBR) mode. 
All material used for the evaluation was fifth-order HOA input and output ($L=36$).
This means the highest test bitrate was $36\cdot48\,\mathrm{kbps}$ = $1728\,\mathrm{kbps}$, with Opus operating on ($J=36$) beamformers/channels.
For this bitrate, we configured a Hoac equivalent case, with $48\,\mathrm{kbps}$ for the $J=12$ TCs (\ie using $12\cdot48\,\mathrm{kbps}$ = $576\,\mathrm{kbps}$), with the remaining bit reserves available for coding the additional metadata. Here, the parameters for the two lowest bands were downsampled by a factor of 2.
For a medium bitrate test-case, we configured the baseline Opus codec to use $768\,\mathrm{kbps}$ in total.
The Hoac equivalent medium bitrate case then used a (slightly non-uniform) $J=9$ channel core coder running at $48\,\mathrm{kbps}$ per channel, with the parameters for the eight lowest bands downsampled by a factor of 2.
For a low bitrate case, we configured Opus to use $512\,\mathrm{kbps}$.
The Hoac equivalent low bitrate case was then to apply a $32\,\mathrm{kbps}$ per channel core coder on $J=6$ TCs, while also down-sampling the eight lowest band parameters by a factor of 2.
It is emphasized that the proposed codec was tuned such that the combined TCs and coded parameters represented slightly less transport data than the baseline Opus. This is because we are aware of potentially missing layers, such as further status bits and error correction bits for the metadata. It is also highlighted that the metadata bitrate is variable by nature, since it depends on the statistics of the scene (as is the case for the VBR mode perceptual audio coder).

Four diverse HOA scenes, identified as representing critical test items in previous studies \cite{hold2024compression,hold2023optimizing}, were included for the formal listening test\footnote{The listening test audio files and further material available here:\\ \url{research.spa.aalto.fi/publications/papers/hoac/}}: 1) a dense \textit{orchestra} simulated with shoebox image-source reverberation; 2) a four-piece funk \textit{band} in a reverberant environment; 3) a scene comprising female speech, clapping, a water fountain, and piano in a reveberant room, using a simulated Eigenmike64 array (\textit{emscene}) encoded into fifth-order HOA; and 4) four \textit{moving} speakers in a reverberant environment, following random trajectories at different and varying velocities. The input fifth-order HOA was then rendered for static headphones playback using the Magnitude Least-Squares (MagLS) Ambisonic binaural rendering \cite{schorkhuber2018}. These uncompressed renders served as the reference cases. The HOA scenes were then encoded and decoded using all combinations of Opus and Hoac at $1728$, $768$ and $512\,\mathrm{kbps}$, and reproduced using the same MagLS Ambisonic rendering. The reference low-pass filtered at $3.5$\,kHz served as an Anchor. In a standard MUSHRA listening test \cite{assembly2015itu}, participants were encouraged to use the looping function within the $10\,\mathrm{s}$ items before making their assessments.

\section{Results and Discussion}
Conventional multi-channel audio codecs operate very efficiently in most typical scenarios.
The issues arise when a codec, such as Opus, is starved of bitrate. For example, the tested $768/36 = 21.3\,\mathrm{kbps}$  and $512/36 = 14.2\,\mathrm{kbps}$ represent severe reductions from the bitrates associated with transparent coding.
While modern codecs may be able to utilize the bit reservoir over channels, which may increase the perceived single channel quality, the limited available average bitrate per channel still poses a challenge.
Therefore, the main proposal of this paper is that instead of audio coding all input channels at such restrictive bitrates, it may be more appropriate to instead reduce the number of TCs and to send additional quantized metadata; corresponding to a perceptually-motivated sound field model.

\begin{figure*}
    \centering
    \includegraphics{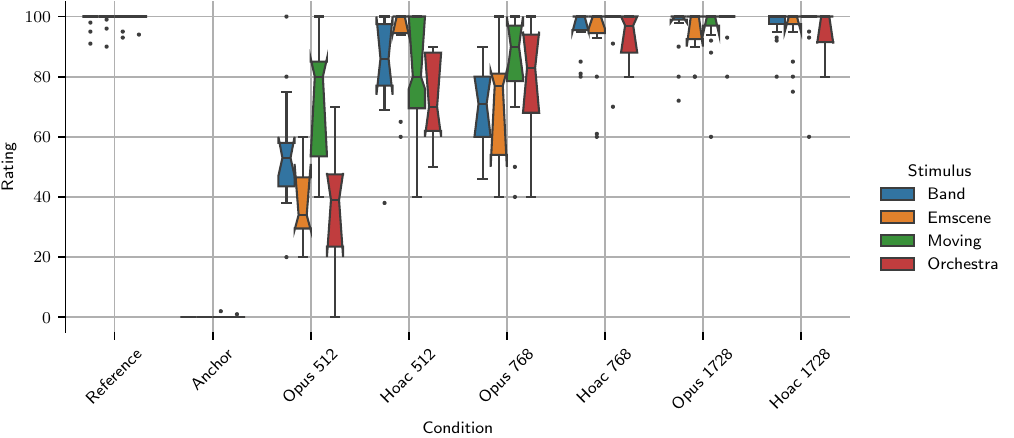}
    \caption{Listening test results plotted independently for each test scene. The boxplots show 95\% confidence interval notches around the median.}
    \label{fig:results-GroupsBox}
    \vspace{-0.1cm}
\end{figure*}

The listening test results are presented in Fig.\,\ref{fig:results-TotalVio} and Fig.\,\ref{fig:results-GroupsBox}, with $95\%$ confidence intervals around the median from bootstrapping ($n=10000$). Participants informally reported that codec distortions, coloration, or shifts in spatial image represented the most prominent artifacts used to identify degredations during the test.
The statistical analysis in the following is reported based upon the non-parametric Wilcoxon signed-rank test (at $\alpha=0.05$) over all responses collected from $17$ subjects, where $2$ were excluded due to rating the reference condition under $90$. 
At $1728\,\mathrm{kbps}$, the proposed (Hoac) and baseline (Opus) codecs saturate towards the end of the rating scale, with medians of $100$. %
The outliers under $100$ were rated consistently between conditions, implying that the degradation may have originated from the core coder (both running at $48\,\mathrm{kbps}$, per channel on average).
At $768\,\mathrm{kbps}$, Opus was rated statistically significantly lower than the reference and Hoac. %
\textit{Hoac\,768} on the other hand was still rated with a median close or equal to $100$, implying that most test participants could not confidently differentiate it from the reference. %
For $512\,\mathrm{kbps}$, we observe larger (statistically significant) differences between Hoac and Opus.
In all but one test case, Hoac was rated higher, with the confidence intervals between \textit{Opus\,512} and \textit{Hoac\,512} not overlapping.
The exception is in item \textit{moving}, with a median similar to Opus, which was rated also fairly high. This indicates that Opus is generally able to code speech sufficiently at these bandwidths (encoder speech detection was enabled), and the degradation in Hoac could here stem from potentially over-simplistic parameter compression, leading to only minor differences between them.

Additional insight is provided by the root-mean-square errors (RMSE) averaged over orders ($\operatorname{mean}(\mathrm{RMS}_\mathrm{n,m}(\mathrm{out}) / \mathrm{RMS}_\mathrm{n,m}(\mathrm{in}))$) and time (cf., companion website). Opus typically introduces little RMSE per channel.
With the exception of \textit{emscene}, the RMSE was under $0.5\mathrm{dB}$ per order for both, Opus and Hoac.
However, based on the conclusions from the listening test, it is clear that observing only such objective metrics is insufficient. 
It is our understanding that restricting the Hoac sound scene model parameter estimations for the low bitrate scenarios, with limited metadata, in turn results in higher errors.
However, especially in the case of $512\,\mathrm{kbps}$, these small shifts in the spatial properties of the sound scene appear to be less disturbing than the coding/distortion artifacts introduced by the bitrate deprived Opus.
This general listener preference is also supported by previous findings judging the quality of spatial audio reproduction \cite{rumsey2005relative}.
One interesting case of note, however, is the simulated microphone array recording (\textit{emscene}). Due to its incorporation of HOA capture limitations, both codecs show increased RMSEs here. %
In Hoac, however, the spatial covariance model forming the target for the post-processing $\vec{g}$ makes idealized assumptions regarding a full expansion order, \ie without decaying components over increasing order, nor modeling array limitations.
Therefore, this leads to an increase in energy in high SH orders compared to the input, which effectively means some degree of sharpening of directional components in the scene.
Future work could therefore include an option to incorporate a more faithful order decay, although Fig.\,\ref{fig:results-GroupsBox} suggests this may have minimal perceptual relevance.

We emphasise that the proposed perceptual audio codec extension was only mildly tuned during the course of this study. However, we believe that this work already highlights the additional freedoms the architecture can facilitate, such as lowering the bitrate of existing codecs in a more perceptually favourable manner. 
For medium bitrates, the results suggest an advantage to spend the bits on fewer, but higher-quality, audio channels, trading in metadata utilization.
Note that in the proposed architecture, increasing bitrate consecutively restores more input channels.
The initial results also suggest robustness to coarse quantization and down-sampling of the utilized parameters.
Furthermore, we highlight that the scene parameterization also enable a much greater scope of user-controllable rendering options, such as upmixing to higher-orders~\cite{hold2023optimizing}, and spatial audio effects \cite{mccormack2021parametric}.

Future work may include a perceptual evaluation of different perceptual audio coders and metadata compression strategies, tuning parameter settings, and also investigating the other perceptually inspired optimizations noted throughout the paper. Besides the presented HO-DirAC model, we postulate that other choices may facilitate parametric spatial audio compression algorithms, such as in~\cite{mccormack2023parametric} were the number of TCs are adapted scene-dependently. %

\section{Conclusion}
In this paper, a spatial audio compression codec utilizing the HO-DirAC parametric model was presented. The codec operates based on down-mixing the HOA input signals via beamforming, where the resulting signals are coded using a perceptual audio coder. The estimated HO-DirAC parameters are grouped over frequency bands, down-sampled, quantized, and Huffman-coded by the encoder. 
Such a codec is built upon an intuitive and perceptually-motivated coding strategy. On the receiving end, the decoder utilizes both the transported audio channels and the coded parameter metadata to reconstruct the HOA audio signals.
Instrumental and perceptual evaluations were conducted for the proposed perceptually-motivated parametric coding strategy, which was compared against an existing HOA audio codec that is based solely on perceptual audio coding. Here, four representative fifth-order HOA scenes were encoded and decoded, with the RMSEs computed and perceived audio quality assessed. It was found that listening test participants rated the proposed parametric spatial audio codec notably higher on average, particularly at low and medium bitrates, with both strategies converging towards the uncompressed reference for higher bitrates.
In conclusion, this suggests that the additional metadata provided by parametric spatial audio models can benefit the perceptual quality of HOA codecs.

\vfill\pagebreak

\bibliographystyle{IEEEbib}
\bibliography{refs}

\end{document}